\documentclass[]{article}
\usepackage{graphicx,amsmath,amssymb,amsbsy,latexsym}
\usepackage[mathscr]{eucal}

\def\R{\mathbb{R}}

\def\N{\mathbb{N}}

\def\tr{\mathrm{tr}}
\def\half{\frac{1}{2}}

\newcommand{\lalg}[1]{\mathfrak{#1}}  
\newcommand{\so}{\mathfrak{so}}
\newcommand{\SO}{\mathrm{SO}}

\newcommand{\mIm}{\mathrm{Im}}

\newcommand{\eqa}{\begin{eqnarray}}
\newcommand{\neqa}{\end{eqnarray}}
\newcommand{\be}{\begin{equation}}
\newcommand{\ee}{\end{equation}}

\newcommand{\Ref}[1]{(\ref{#1})}

\newcommand{\scrM}{\mathcal{M}}

\newcommand{\scrI}{\mathcal{I}}
\newcommand{\scrT}{\mathcal{T}}

\newcommand{\dual}{\,\,{}^\star\!}

\newcommand{\pb}[1]{\underleftarrow{#1}}

\newcommand{\Hil}{\mathcal{H}}

\newcommand{\Link}{\mathrm{Link}}

\newcommand{\mspan}{\mathrm{span}}  

\newcommand{\dif}{\mathrm{d}}

\newcommand{\Gfactor}{\kappa}

\begin{document}

\title{Coherent states, constraint classes, and area operators in the new
spin-foam models\\[.5mm]}
\author{Jonathan Engle, Roberto Pereira
 \\[1mm]
\normalsize \em CPT%
\footnote{Unit\'e mixte de recherche (UMR 6207) du CNRS et des Universit\'es
de Provence (Aix-Marseille I), de la Meditarran\'ee (Aix-Marseille II) et du Sud (Toulon-Var); laboratoire affili\'e \`a la FRUMAM (FR 2291).} , CNRS Case 907, Universit\'e de la M\'editerran\'ee, F-13288 Marseille, EU
}
\date{\today}

\maketitle\vspace{-7mm}

\begin{abstract}\noindent

Recently, two new spin-foam models have appeared in the literature,
both motivated by a desire to modify the Barrett-Crane model in such
a way that the imposition of certain second class constraints,
called cross-simplicity constraints, are weakened. We refer to these
two models as the FKLS model, and the flipped model.
Both of these models are based on a reformulation of the
cross-simplicity constraints. This paper has two main parts.
First, we clarify the structure of the reformulated cross-simplicity
constraints and the nature of their quantum imposition in the new
models. In particular we show that in the FKLS model, quantum
cross-simplicity implies no restriction on states. The deeper reason
for this is that, with the symplectic structure relevant
for FKLS, the reformulated simplicity constraints, among themselves,
now form a \textit{first class} system, and this causes the coherent
state method of imposing the constraints, key in the FKLS model, to
fail to give any restriction on states. Nevertheless, the
cross-simplicity can still be seen as implemented via suppression of
intertwiner degrees of freedom in the dynamical propagation. In the
second part of the paper, we investigate area spectra in the models.
The results of these two investigations will highlight how, in the
flipped model, the Hilbert space of states, as well as the spectra
of area operators exactly match those of loop quantum gravity,
whereas in the FKLS (and Barrett-Crane) models, the boundary Hilbert
spaces and area spectra are different.

\end{abstract}


\section{Introduction}

Loop quantum gravity (LQG) is a modern, background independent
approach to the canonical quantization of general relativity.  For
reviews, see \cite{rovellibook, alrev, thiemannbook}. The kinematics
of LQG are well-understood, whereas the dynamics is less
well-understood. One approach to the dynamics is the canonical
approach \cite{thomas}. A second approach, which seeks to preserve
manifest space-time covariance, is a sum-over-histories approach,
leading to the spin-foam formalism \cite{spinfoams}. In the search
for a spin-foam model of quantum gravity, recent progress has been
made in better understanding how to handle certain second class
constraints
--- called the \textit{simplicity} constraints. In the most
prominent spin-foam model, the Barrett-Crane model \cite{bc}, the
simplicity constraints are imposed strongly as operator equations.
Because of this, all intertwiner degrees of freedom in the spin-foam
model are frozen out; this has caused problems with the
semiclassical limit of the theory, as investigated in
\cite{emanuele}. In response to this problem, it was realized that
the simplicity constraints should be handled more carefully, as they
are second class, with the hope that the necessary intertwiner
degrees of freedom would be liberated. From this motivation, two
alternatives to the Barrett-Crane model have recently been proposed
\cite{eprlett, eprpap, ls_cohstates, fk2007, ls_model}. These two
alternative models can be viewed as corresponding to the case of
small Barbero-Immirzi parameter $\gamma$, and the case of infinite
Barbero-Immirzi parameter\footnote{In this paper, we do not
construct the quantum theories for arbitrary $\gamma$ and then
restrict to $\gamma \ll 1$ and $\gamma = \infty$. Rather, we let the
assumption of $\gamma \ll 1$ and $\gamma = \infty$ influence the
quantizations of the constraints, as well as the quantizations of
the area operators. Nevertheless, in the forthcoming paper
\cite{gengamma} a family of quantum theories will be constructed for
finite $\gamma$, and the flipped model will be seen to correspond to the
$\gamma \ll 1$ case.  Furthermore, the FKLS model, in addition to corresponding
to the $\gamma = \infty$ case, arises as the $\gamma \rightarrow \infty$
limit of another family of quantum theories for finite $\gamma$, different
from the family to be introduced in \cite{gengamma} (see \cite{fk2007}).
However, the BC model, as far as we know, corresponds only to $\gamma = \infty$,
and not to any limit within a family of theories for finite $\gamma$.}.
The value of $\gamma$ affects only the symplectic
structure of the canonical theory.  For
small $\gamma$ and infinite $\gamma$, using the terminology of \cite{eprlett,
eprpap}, this symplectic structure is, respectively, the ``flipped''
or ``unflipped'' symplectic structure. We will therefore refer to
the small $\gamma$ model \cite{eprlett, eprpap} as the ``flipped
vertex.'' The infinite $\gamma$ model, as quantized in \cite{fk2007},
will be referred to as the FKLS (Freidel-Krasnov-Livine-Speziale)
model.

Both of these models can be obtained by using the coherent state
approach to solving the cross-simplicity constraints developed in
\cite{ls_cohstates, ls_model, fk2007}.  In addition, the flipped
model \cite{eprlett, eprpap} can be obtained using an approach to
the constraints involving Casimir operators; as we will see in this
paper, this can be viewed as a sort of master constraint
\cite{master} approach to solving the simplicity
constraints. This is the original way in which the flipped
model was derived. The fact that the model was later derived using
coherent states was complete surprise at the time and greatly
increased confidence in the model, as well as opening a possible
avenue for investigating its semiclassical limit.

We will present a clarification of the constraint analysis involved
in these two models.  As a consequence, we will call into question
the manner of imposing the constraints in the FKLS model.  More
specifically, we will note that if one tries to interpret the
imposition of constraints in FKLS as an imposition of constraints on
states, in fact, FKLS does not impose cross-simplicity at all -- the
$SO(4)$ intertwiner spaces remain completely
unconstrained\footnote{By ``constraints on states'' we mean first
and foremost constraints as imposed in the boundary Hilbert space of
the model.  In \cite{fk2007, ls_model}, the boundary Hilbert space of
FKLS is not discussed; in \S\ref{qkin}, \S\ref{cross_sect} of 
this paper, we derive it in what seems to us the most straightforward way.}. 
The source of this will be found to be a curious property of the
reformulation of the constraints at the heart of the two new
spin-foam models: in the unflipped case the cross-simplicity
constraints form a closed, first class algebra, whereas in the
flipped case they do not. This is relevant for the following reason.
The procedure of imposing the constraints in the new models
\cite{eprlett, eprpap, ls_model, fk2007} consists in two steps:
first impose the simplicity constraints and then average over
$SO(4)$ gauge-transformations. It is in the first step that one uses
the coherent state approach. However, as we will show through a
simple example, it seems that quite generally a first class
constraint system cannot be imposed using coherent states: such a
procedure seems to result in no constraints being imposed on the
states, and this is what underlies the apparent difficulty with
imposition of cross-simplicity in the FKLS 
model. (Thus, note that we are not questioning the coherent state method 
as such, but only its application in FKLS.)

Rather, with the reformulated constraints, in the case of the unflipped
symplectic structure, because the cross-simplicity constraints
now form a first class system, they should be imposed 
strongly in the first step mentioned above.  But this leads to the 
Barrett-Crane model.

The ``master constraint'' approach to cross-simplicity allows one
to be less concerned with the class of the constraints involved,
as the master constraint
program applies to any type of constraints, whether first or second
class \cite{master}.  Using the master constraint
approach (which, as we will see, is the same as the ``Casimir operator''
approach of \cite{eprlett, eprpap}), for the unflipped symplectic
structure, we obtain Barrett-Crane, whereas for the flipped symplectic
structure, we obtain the flipped model of
\cite{eprlett, eprpap}.

One result of this investigation is that, in the FKLS model, as the
$SO(4)$ intertwiner spaces are completely unconstrained, the boundary
space is clearly not isomorphic to the Hilbert space of LQG, in
contrast with the flipped model \cite{eprlett, eprpap}. In addition,
we will close with a discussion of $SO(3)$-gauge-fixed area
operators. We will note that in the flipped model, the spectrum of
this operator exactly matches that of the area operator in LQG,
including numerical factors, whereas the spectra in the FKLS model
and BC models are different.

A final note should be said regarding the FKLS model.
Although the FKLS model
does not impose cross-simplicity as a constraint on states,
nevertheless it imposes cross-simplicity in a different,
albeit less standard sense: the dynamics appears to suppress intertwiners
that are far from the Barrett-Crane intertwiner.  This
viewpoint will be touched upon in the discussion section of
the paper.

The paper is organized as follows.  First, we will briefly review
the structure of the classical discrete theory from
\cite{eprlett, eprpap}. Then we will discuss the space of states
satisfying cross-simplicity in the BC, flipped, and FKLS models.
This will in part motivate a subsequent section discussing the
relation between the coherent state method of imposing constraints,
and the class of the constraints involved. Lastly,
the area spectra are analyzed in the BC, flipped, and FKLS models
and are compared with the spectra in LQG.  Some final reflections on
the significance of these results are then given.

\section{Model and constraints}

\subsection{Discrete classical theory}
\label{modelcl_sect}

Following \cite{eprpap}, we introduce a Regge geometry. That is,
first we introduce a simplicial decomposition $\Delta$ of
space-time, consisting of 4-simplicies, tetrahedra, and triangles.
These are dual respectively to verticies, edges, and faces in the
dual 2-complex, and we shall denote them by $v$, $t$, and $f$.
Geometry is flat on each 4-simplex. Curvature is concentrated on the
``bones'' $f$, and is coded in the holonomy around the ``link'' of
each $f$.

The basic discrete variables for the theory
can be motivated as follows.  First, for each 4-simplex $v$,
introduce a tetrad field $e_\mu^I(v)$, defined within $v$, and
for each tetrahedron $t$, introduce a tetrad field $e_\mu^I(t)$,
defined within the two 4-simplicies adjoining $t$.  We require that
all of these tetrads determine the same, locally flat geometry
where they overlap, and we require that they all be covariantly
constant with respect to the derivative operator
determined by this geometry.
%
%
For each $t$ and triangle $f$
therein, we then define $B_f(t) \in \so(4)$ by
\begin{equation}
\label{bdef}
B_f(t)^{IJ} := \dual \int_{f} e^I(t) \wedge e^J(t)
= \frac{1}{2}\epsilon^{IJ}{}_{KL} \int_f e^K(t) \wedge e^L(t),
\end{equation}
and for each 4-simplex $v$ and tetrahedron $t$ therein, we define
$(V_{vt}^{-1}\equiv) V_{tv} \in \SO(4)$ by
\begin{equation}
\label{vdef}
e_\mu^I(t) = (V_{tv})^I{}_J e_\mu^J(v).
\end{equation}
(\ref{bdef}) and (\ref{vdef}) are the basic discrete space-time
variables. For each triangle $f$ and each pair of tetrahedra $t,t'$
in the link of $f$, define
\begin{equation}
U_f(t,t') := V_{tv_1} V_{v_1 t_1} V_{t_1 v_2} \cdots V_{v_n t'}
\end{equation}
where the product is around the link in the clock-wise direction
from $t'$ to $t$.
The constraints on the discrete variables may then be stated as follows.
\begin{enumerate}
\item $U_f(t,t') B_f(t') = B_f(t) U_f(t,t')$ \qquad $\forall$ $f$ and
$t,t' \in \Link(f)$

\item (closure) $\sum_{f \in  t} B_f(t) = 0$ \qquad $\forall$ $t$

\item discrete simplicity constraints
\begin{displaymath}
\begin{array}{rrccll}
\nonumber
\text{(i)}&C_{ff}:=&
\dual B_f(t)\cdot B_f(t)
&\approx& 0 &\qquad \forall f\\
\nonumber
\text{(ii)}&C_{ff'}:=&
\dual B_f(t)\cdot B_{f'}(t)
&\approx& 0 &\qquad \forall f,f' \in t\\
\nonumber
\text{(iii)}&&\dual B_f(v)\cdot B_{f'}(v)
&\approx& \pm 12V(v) &\qquad \forall f,f' \in v\text{ not in the same }t
\end{array}
\end{displaymath}
\end{enumerate}
where $\dual$ stand for the Hodge dual in the $SO(4)$ algebra indices and the dot stands for the scalar product in this algebra.
This is the traditional formulation of the constraints.
The two constraints (3.i) and (3.ii) in fact have two sectors
of solutions, one in which $B = \dual e \wedge e$, and one
in which $B = e \wedge e$. For finite, non-trivial
Barbero-Immirzi parameter, both sectors in fact yield
GR, but the value of the Newton
constant and Barbero-Immirzi
parameter are different in each
sector\footnote{In
the $B=\dual e \wedge e$ sector, with coefficients in the action
as in (\ref{discaction}), one obtains the Holst \cite{holst} formulation of
general relativity with Newton constant $G$ and Barbero-Immirzi
parameter $\gamma$.  In the sector $B=e \wedge e$, one \textit{also}
obtains the Holst formulation of general relativity, but
this time with $G\gamma$ acting as the Newton constant, and
$\frac{1}{\gamma}$ acting as the Barbero-Immirzi parameter.}.
Therefore, in order to talk about one sector at a time in
a coherent way, it is desireable to reformulate the
simplicity constraints such that these two sectors are distinguished.
In fact, this can be done: condition (3.ii) can be replaced with
the condition that
\begin{center}
\begin{tabular}[t]{rl}
(3.ii')& For each tetrahedron $t$, \textit{there exists}
an internal vector $n^I$ such that \\
& $(\dual B_f(t))^{IJ}n_I = 0$ for all $f \in t$.
\end{tabular}
\end{center}
This reformulation of the constraint (3.ii)
(the ``off-diagonal'', or ``cross-simplicity'' constraint)
is central to the new models \cite{eprlett, eprpap, fk2007, ls_model}.
When constructing the quantum theory, the above constraints
are incorporated as follows.
(1.) will be imposed prior
to varying the action.
(2.),(3.i),(3.ii') will be imposed in quantum theory.
As noted in
\cite{eprpap}, (3.iii) is automatically satisfied
when the rest of the constraints
are satisfied, due to the choice of variables.

The classical discrete action is
\cite{eprpap}
\begin{eqnarray}
\nonumber
S_{disc.} &=& -\frac{1}{2\Gfactor} \sum_{f \in int\Delta} \tr\left[B_f(t)U_f(t)
+ \frac{1}{\gamma}\dual B_f(t) U_f(t) \right]  \\
\label{discaction}
&& -\frac{1}{2\Gfactor} \sum_{f \in \partial\Delta} \tr\left[B_f(t)U_f(t,t')
+ \frac{1}{\gamma}\dual B_f(t) U_f(t,t') \right]
\end{eqnarray}
where $U_f(t):= U_f(t,t)$ is the holonomy around the full link,
starting at $t$, and where we have set $\Gfactor = 8\pi G$.
%
%
%
%
From this we can read off the boundary variables as
$B_f(t) \in \mathfrak{so}(4)$, $U_f(t,t') \in SO(4)$.
One can also see that the variable conjugate to $U_f(t,t')$ is
proportional to $B_f(t) + \frac{1}{\gamma} \dual B_f(t)$.
The constant of proportionality is fixed in appendix \ref{scale_app}:
\begin{equation}
\label{j_interms_b}
J_f(t) = \frac{1}{\Gfactor}\left(B_f(t) + \frac{1}{\gamma}\dual B_f(t)\right).
\end{equation}
More precisely, each matrix component $J_f(t)^{IJ}$ has as its
Hamiltonian vector field the left invariant vector field on the
group $U_f(t,t')$ corresponding to the lie algebra element $J^{IJ}$
defined in appendix \ref{alg_app}. Inverting the above equation
gives
\begin{equation}
\label{b_interms_j}
B_f(t) := \Gfactor \left(\frac{\gamma^2}{\gamma^2-1}\right)
\left(J_f(t)-\frac{1}{\gamma}\dual J_f(t)\right)
\end{equation}
For the cases $\gamma \ll 1$ and $\gamma = \infty$,
this reduces to
\begin{align}
\nonumber \gamma \ll 1 &\;\,\,\rightarrow\;\,\,
B = \Gfactor \gamma \dual J &\qquad \gamma = \infty
&\;\,\,\rightarrow\;\,\, B = \Gfactor J .
%
%
\end{align}
corresponding respectively to the flipped and non flipped Poisson
structures of $SO(4)$. In terms of the new variables $J_f$ the
constraints (3.i) and (3.ii') read \footnote{A continuum version of
the first of these equations appeared in \cite{eteraBF}}(we consider
$\gamma \neq 0,1$ so that overall factors can be discarded):
%
%
\begin{eqnarray}
C_{ff}&:=&\dual J_f\cdot J_f\left(1+\frac{1}{\gamma^2}\right)-\frac{2}{\gamma}J_f\cdot J_f \approx 0 \label{Cf}\\
\label{crosssimp} C_f^J&:=&n_I\left((\dual
J_f)^{IJ}-\frac{1}{\gamma}J_f^{IJ}\right) \approx 0 \label{CJ}
\end{eqnarray}
The closure for the $B_f$ is equivalent to the closure for the $J_f$
which, as noted in \cite{eprpap}, will be imposed automatically by
the dynamics. The strategy for imposing (\ref{crosssimp}) in quantum
theory will be to first impose it for a fixed $n^I$, and then to
average over $SO(4)$. This is the strategy used in
\cite{eprlett,eprpap,fk2007,ls_model}. For this purpose, let us fix
$n^I := \delta_{I0}$ in what follows. Equation \Ref{CJ} then becomes
\begin{equation}
\label{GF_3ii}
C_f^j=\frac{1}{2}\epsilon^j{}_{kl}J_f^{kl}-\frac{1}{\gamma}J^{0j}=L_f^j-\frac{1}{\gamma}K_f^j
\end{equation}
where $\epsilon^{j}{}_{kl}:=\epsilon^{0j}{}_{kl}$,
$L_f^j:=\frac{1}{2}\epsilon^{j}{}_{kl}J^{kl}$, $K_f^j:=J^{0j}$. We
further make the self-dual/anti-self-dual decomposition of
$J_f^{IJ}$:
\be
\label{sddecomp}
J_f^{(\pm)i}:=\frac{1}{2}\left(\pm K_f^i +
L_f^i\right)
\ee
so that $L_f^i=J_f^{(+)i}+J_f^{(-)i}$ and
$K_f^i=J_f^{(+)i}-J_f^{(-)i}$. In terms of this decomposition we
rewrite the constraints as follows:
\begin{eqnarray}
C_{ff}&=&\left(J_f^{(+)}\right)^2\left(1-\frac{1}{\gamma}\right)^2-\left(J_f^{(-)}\right)^2\left(1+\frac{1}{\gamma}\right)^2 \approx 0 \label{C} \\
\label{GFSD_3ii}
C_f^j&=&J_f^{(+)j}\left(1-\frac{1}{\gamma}\right)+J_f^{(-)j}\left(1+\frac{1}{\gamma}\right)
\approx 0 \label{Cj}
\end{eqnarray}

We take (\ref{C},\ref{Cj}) as our basic set of constraints. They
will be imposed in quantum theory.  Because (\ref{GFSD_3ii}) is a
gauge-fixed version of the constraint (3.ii'), one will have to
average over $SO(4)$ gauge transformations after its
imposition.\footnote{\label{timegaugefn}One can be more precise
about the gauge used in \Ref{GFSD_3ii}: it is just the usual time
gauge, used, for example, in LQG.  More precisely, it is implied by
the usual time gauge. For, suppose one is given a spatial
hypersurface $\Sigma$ and a 4-vector field $t^a$ thereon, transverse
to $\Sigma$, and consider the fixed internal vector $n^I =
\delta^I_0$. The corresponding time gauge implies that $n_I
e^I_{\pb{a}} = 0$ where the underarrow denotes pull-back to
$\Sigma$.  This in turn implies that for any triangle $f$ in
$\Sigma$, $n_I \dual B^{IJ}_f := n_I \int_f e^I \wedge e^J = 0$.}

Let us note something which will be important later. For $\gamma = \infty$
(the unflipped symplectic structure), the gauge-fixed cross-simplicity constraints (\ref{GF_3ii})
become
\begin{equation}
\label{bigg_3ii}
C^j_f = L^i_f = J^{(+)i}_f + J^{(-)i}_f = L^i_f
\end{equation}
and for $\gamma \ll 1$ (the flipped symplectic structure) they become
\begin{equation}
\label{smallg_3ii}
C^j_f \propto -K^i_f = -J^{(+)i}_f + J^{(-)i}_f .
\end{equation}
In the unflipped case, the constraints (\ref{bigg_3ii}) close
and so are ``first class'' in this sense.  In fact their algebra is
just that of $\so(3)$.  In the flipped case, the constraints
(\ref{smallg_3ii}) do not close and so are ``second class'' in this
sense.

\subsection{Quantum kinematics}
\label{qkin}

From the discrete boundary variables and their symplectic
structure, one can write down the Hilbert space associated
with a boundary or 3-slice.
To do this, it is simpler to
switch to the dual, 2-complex picture, $\Delta^*$.
For each 3-surface $\Sigma$ intersecting no vertices of $\Delta^*$,
let $\gamma_{\Sigma} := \Sigma \cap \Delta^*$.  The Hilbert space
associated with $\Sigma$ is then
\begin{equation}
\Hil_{\Sigma} = L^2\left(SO(4)^{|L(\gamma_\Sigma)|}\right)
\end{equation}
where $|L(\gamma_\Sigma)|$ denotes the number of links in
$\gamma_\Sigma$. Let $\hat{J}_f(t)^{IJ}$ denote $(-i)$ times the
left-invariant vector fields, determined by the basis $J^{IJ}$ of
$\mathfrak{so}(4)$, on the copy of $SO(4)$ associated with the link
$l = f \cap \Sigma$ determined by $f$, with orientation such that
the node $n = t \cap \Sigma$ is the source of $l$. The $B_f(t)$'s
are then quantized as
\begin{equation}
\label{Bquant}
\hat{B}_f(t) := \Gfactor \left(\frac{\gamma^2}{\gamma^2-1}\right)
\left(\hat{J}_f(t)-\frac{1}{\gamma}\dual \hat{J}_f(t)\right)
\end{equation}

Next we promote \Ref{C} and \Ref{Cj} to quantum operators. We note
that the first constraint commutes with the others and can be
carried directly to quantum theory. In terms of the usual
(generalized) spin-network basis
$\Psi_{(\vec{j}^+,\vec{j}^-),\vec{T}}$ labelled by spins
$(j_f^+,j_f^-)$ on links and tensors $T_t$ at nodes, the first
constraint implies
\begin{equation}
j_f^+ (j_f^+ + 1)=\left(\frac{1+\gamma}{1-\gamma}\right)^2\;j_f^- (j_f^- + 1)
\end{equation}
For either $\gamma \ll 1$ or $\gamma = \infty$
this condition is
satisfied by the simple representations of $SO(4)$, i.e., $j^+=j^-$.
In the following
we will always specialize to either $\gamma \ll 1$ or $\gamma = \infty$.

\section{Cross-simplicity in the various models, and the solution spaces}
\label{cross_sect}

Up until now, the quantum theory is standard, and the same as in the
BC model.  The difference among the BC model, the flipped model, and
the FKLS model comes in the implementation of the cross-simplicity
constraints. Whereas the diagonal simplicity constraints constrain
the spins on the $SO(4)$ spin-networks, the cross-simplicity
constraints constrain the intertwiners.  For simplicity of
presentation, in this section we will consider a fixed node $n$ and
the intertwiner space at that node.  We number the links at this
node $1,2,3,4$.  For external spins $(j_1, j_1), \dots, (j_4,j_4)$,
the space of possible tensors at the node $n$ will be denoted
\begin{equation}
\label{tensp_def} \scrT^{(\vec{j},\vec{j})}:= \Hil_{(j_1,j_1)}
\otimes \cdots \otimes \Hil_{(j_4,j_4)} = \left(\Hil_{j_1} \otimes
\Hil_{j_1}\right) \otimes \cdots \otimes \left(\Hil_{j_4} \otimes
\Hil_{j_4}\right)
\end{equation}
where $\Hil_{j}$ denotes the carrying space for the spin $j$
irreducible representation (irrep) of $SU(2)$ and $\Hil_{(j^+,j^-)}$
denotes the carrying space for the irrep of $SO(4)$ labeled by the
spins $(j^+,j^-)$.  The associated intertwiner subspace at $n$ will
be denoted $\scrI^{(\vec{j},\vec{j})} \subset
\scrT^{(\vec{j},\vec{j})}$.

In the following we will discuss the solutions to cross-simplicity
in the various models \textit{prior} to averaging over $SO(4)$ gauge
transformations, for simplicity of comparison.  That is, we will
discuss imposing the ``gauge-fixed'' cross-simplicity contraint
(\ref{GF_3ii}) in the various models. In each model, this will give us a
``gauge-fixed'' solution subspace of $\scrT^{(\vec{j},\vec{j})}$,
which, when averaged over $SO(4)$ gauge transformations will yield
the final physical space of intertwiners to be summed over in the
spin-foam sum, and to be used in building the physical boundary
Hilbert space.

\subsection{BC model}

In the BC model, we take the unflipped symplectic structure
--- that is, $\gamma = \infty$. As noted above, the
\textit{new, reformulated} cross-simplicity constraints
(\ref{GF_3ii}) in this case close, forming a first class system.
Thus, the contraints \Ref{Cj} can all be imposed simultaneously as
operator equations. We have $C^i_f\sim L_f^i\approx 0$, which
restricts the intertwiner space to be one dimensional, being spanned
by the one unique Barrett-Crane intertwiner \cite{eteraprojsn, reisenberger}.

\subsection{The new spin-foam models}
\label{newmodels}

Both of the new spin-foam models can be viewed as arising from the
use of coherent states to impose the cross-simplicity constraints.
The idea of using coherent states to impose second class constraints
can be found, for example, in \cite{engle2006} (or in more implicit
form in \cite{asht_tate}). The specific coherent states relevant for
the new spin-foam models were introduced by Livine and Speziale in
\cite{ls_cohstates}. We here review these coherent states (see
\cite{ls_cohstates} and appendix \ref{app_coh} for further details).

Consider the $SU(2)$ coherent states $|j,\hat{n}\rangle$, where
$j\in\mathbb{N}/2$ and $\hat{n}$ is a unit vector in $\mathbb{R}^3$.
$|j,\hat{n}\rangle$ may be defined as the $m=j$ eigenstate of
$\hat{n} \cdot \hat{J}$, where $\hat{J}^i := \frac{i}{2} \sigma^i$
is a basis of the Lie algebra satisfying the usual angular momentum
commutation relations. Starting with these, one can construct
$SO(4)$ coherent states by tensoring them together giving states of
the form $|j^+,\hat{n}^+\rangle\otimes|j^-,\hat{n}^-\rangle$. On
such states, the expectation value of the constraint \Ref{GFSD_3ii} is
given by
\be
\langle\vec{C}_f\rangle=j^+\left(1-\frac{1}{\gamma}\right)\hat{n}^+
+ j^-\left(1+\frac{1}{\gamma}\right)\hat{n}^-\approx 0. \ee One can
see that $\hat{n}^+=\hat{n}^-$ and $\hat{n}^+=-\hat{n}^-$ are
solutions for $\gamma \ll 1$ and $\gamma =  \infty$ respectively. The
first set of states define the flipped model \cite{eprlett,eprpap}
as shown in \cite{ls_model}\footnote{The original derivation in
\cite{eprlett,eprpap} did not use coherent states, but rather a
``master constraint''-like approach, which is reviewed in subsection
\ref{flippedsoln} below.  We here use coherent states for unity
of initial presentation of the two new models.}. The second set of states
define the FKLS model (\cite{fk2007},\cite{ls_model}).  In the first
case the constraints \Ref{GF_3ii} are actually satisfied as matrix
elements: that is, given any two $SO(4)$ coherent states
$|\psi\rangle$ and $|\chi\rangle$ constrained by
$\hat{n}^+=\hat{n}^-$, we have $\langle\chi |\; C_f^i\;
|\psi\rangle\approx 0$. This fact is also true for the original form
of the constraints $C_{ff'}$ as was noted in \cite{eprlett, eprpap}.

Up to now we have been considering only coherent states in the
carrying space of a representation associated to a single face, but
we could as well consider intertwiner spaces between the four faces
meeting in a single tetrahedron. All we have to do to get a state in
this space is to tensor four coherent states and project the
resulting state into the gauge invariant subspace. Let us state this
in equations. It will be convenient to
first define the gauge-fixed tensors in the two models:
\begin{eqnarray}
\label{fl_int_gf}
I_{n}^{+, GF}(\hat{n}_a)&=& \otimes_{a=1}^4\; \left(|j_a,\hat{n}_a\rangle\otimes |j_a,\hat{n}_a\rangle\right) \\
\label{fkls_int_gf} I_{n}^{-, GF}(\hat{n}_a)&=& \otimes_{a=1}^4\;
\left(|j_a,\hat{n}_a\rangle\otimes |j_a,-\hat{n}_a\rangle\right)
\end{eqnarray}
so that the $+$ label corresponds to the flipped model and the $-$
label corresponds to FKLS (the notation being motivated by the sign
of the second $\hat{n}$ on the right hand side). Projection of these
to the $SO(4)$ gauge-invariant subspace then yields
\begin{eqnarray}
\label{fl_int} I_{n}^{+}(\hat{n}_a)&=&\int_{SO(4)}\; dG\;
G\cdot I_{n}^{+, GF}(\hat{n}_a) \\
\label{unfl_int} I_{n}^{-}(\hat{n}_a)&=&\int_{SO(4)}\; dG\; \;G\cdot
I_{n}^{-, GF}(\hat{n}_a)
\end{eqnarray}
Transforming to the basis $|j_a;i^+,i^-\rangle$, one can check that these intertwiners can be written as
\begin{eqnarray}
\label{fl_int_sdasd}
I_{n}^{+}(\hat{n}_a)&=&\sum_{i^+,i^-}c_{i^+}(\hat{n}_a)c_{i^-}(\hat{n}_a)|j_a;i^+,i^-\rangle \\
\label{unfl_int_sdasd}
I_{n}^{-}(\hat{n}_a)&=&\sum_{i^+,i^-}c_{i^+}(\hat{n}_a)\bar{c}_{i^-}(\hat{n}_a)|j_a;i^+,i^-\rangle
\end{eqnarray}
where the coefficients $c_i(\hat{n}_a)$ are given in appendix \ref{app_coh}.
The intertwiner states in the flipped model are symmetric under parity
operation (that is, under interchange of self-dual and
anti-self-dual components) and satisfy the $C_{ff'}$ as matrix
elements. Intertwiners in the FKLS model are complex conjugated
under interchange of self-dual and anti-self-dual parts, and satisfy the
constraints only as expectation values.

Equations (\ref{fl_int}) and (\ref{unfl_int}) specify a restriction
to certain $SO(4)$ coherent states in each model.  What is the
corresponding space of solutions to the cross-simplicity
constraints?  If the coherent states are to be considered solutions,
and the solution space is to be a vector space, one has to consider
the span of all the constrained coherent states as the solution space.  We will
look at these spans in the case of the flipped model and the FKLS
model and see what they are; in the FKLS case we will see that the
span is in fact the entire space of $SO(4)$ intertwiners at the node.
These constrained spaces of intertwiners for the models will then be the
intertwiners one sums over in the spin-foam sum, as well as the spaces
of intertwiners used in describing the boundary state space for each model.

For the purposes of describing these solution spaces, for each of
the four links $a=1,2,3,4$ at the node of interest, let
$\hat{L}_a^i$ denote the rotation generators introduced earlier,
generating the $SO(3)$ subgroup preserving the fixed vector $n^I$.
For each $a$, $\hat{L}_a^i \hat{L}_{ai}$ is then the Casimir
operator for the representation of this subgroup on each of the four
links.  For a given set of fixed external spins $(j_1,j_1), \dots
(j_4,j_4)$, the spectrum of each of these Casimir operators is
$\{k_a(k_a+1)\}$ with $k_a \in \{0,1,\dots,2j_a\}$.  One then has a
decomposition of the tensor space at the node into the simultaneous
eigenspaces of these $SO(3)$ Casimirs:
\begin{equation}
\label{decompSO3} \scrT^{(\vec{j},\vec{j})} = \otimes_{a=1}^4
\oplus_{k_a=0}^{2j_a} \Hil_{k_a}
\end{equation}
That is, this is the decomposition of the tensor space into
irreducible representations of the $SO(3)$ subgroup preserving the
chosen gauge-fixed $n^I$.

\subsection{The solution space for the flipped model}
\label{flippedsoln}

At the gauge-fixed level, the solution space for the flipped model
is easy to state. In terms of the decomposition (\ref{decompSO3}),
it is the $k_a = 2j_a$ subspace:
\begin{equation}
\Hil^{Flipped, GF}_{n} = \otimes_{a=1}^4 \Hil_{2j_a} \subset
\scrT^{(\vec{j},\vec{j})}
\end{equation}
See \cite{fk2007, ls_model} for proof that this is the space spanned
by (\ref{fl_int_gf}) (so that its projection to the gauge-invariant
subspace is spanned by \Ref{fl_int}). As noted above, the constraint operators corresponding to
\Ref{smallg_3ii} (as well as the operators corresponding ot the original
formulation of cross-simplicity in (3.ii) above) have zero matrix elements
on this space.  Furthermore, this space can also be obtained by imposing
a sort of ``master constraint''
\cite{master}
constructed from the gauge-fixed
constraints \Ref{GFSD_3ii}.  As above, label the links at the node $n$
by $a\in\{1,2,3,4\}$, and fix the
spins $j_a$ on each of these links.  For each link, labelled by
$a \in \{1,2,3,4\}$, one can then define a ``master constraint''
\begin{equation}
\hat{M}_a = \hat{C}^i_a \hat{C}_{ai}
\end{equation}
acting on the space $\scrT^{(\vec{j},\vec{j})}$.
Each operator $\hat{M}_a$ has as minimal
eigenvalue $2j_a\hbar^2$
%
%
(a value zero in the semiclassical limit\footnote{ The semiclassical
limit is the limit $\hbar \rightarrow 0$ while holding the analogues
of classical quantities fixed. With diagonal simplicity satisfied,
as here, the eigenvalues $A_4^2$ of the full $SO(4)$ area squared
are proportional to $\hbar^2 j(j+1)$. Thus, taking $\hbar
\rightarrow 0$ holding $A_4$ fixed, $\hbar^2 j^2$ approaches a
constant, so that $\hbar^2 j$ goes to zero.}).  Just to better
follow the prescription of \cite{master}, let us sum these four
master constraints into a single constraint for the node:
\begin{equation}
\hat{M}_n := \sum_{a=1}^4 \hat{M}_a
\end{equation}
As all the constraints $\hat{M}_a$ commute with each other, the
spectrum of $\hat{M}_n$ will just be the point-wise sum of the
spectra of the operators $\hat{M}_a$, so that the minimal eigenvalue
will be $\sum_{i=1}^4 2 j_a \hbar^2$. Thus, following the
prescription of \cite{master}, we take the solution space to be the
eigenspace of $\hat{M}_n$ with minimal eigenvalue $\sum_{i=1}^4
2j_a\hbar^2$. One can check that this space is precisely
$\Hil^{Flipped, GF}_{n}$. (This presentation is a different way of
looking at the original derivation of $\Hil^{Flipped, GF}_{n}$ in
\cite{eprlett, eprpap}.)

\subsection{The solution space for the FKLS model}

In this subsection we wish to understand more explicitly the
intertwiner solution space for the FKLS model:
\begin{equation}
\Hil^{FKLS, \; GF}_{n} := \mspan\left\{I_{n}^{-,GF}(\hat{n}_a)
\right\}_{\hat{n}_a}.
\end{equation}
For this purpose, let us first define, for each $j \in \N/2$,
\begin{equation}
G_{j} := d_j^2 \int_{g \in SU(2)} \dif g |j, g \cdot \hat{n}_o
\rangle \otimes |j, -g \cdot \hat{n}_o \rangle \otimes \langle j, g
\cdot \hat{n}_o | \otimes \langle j, - g \cdot \hat{n}_o |
\end{equation}
so that $G_{j}$ maps $\Hil_{(j,j)}$ to
$\Hil_{(j,j)}$. Here, $d_j := 2j+1$ is a
dimension factor, and $\hat{n}_o$ is an arbitrary unit reference vector
in $\R^3$. Because of the integration over the action of $g$, the
above expression is independent of the choice of $\hat{n}_o$. $G_j$
so defined is in fact the same as the $G_j$ defined in equation (76)
in \cite{fk2007}.  Next, for a fixed node $n$ with external spins,
$(j_1, j_1), \dots (j_4,j_4)$, consider the map
\begin{equation}
\label{nodeG_def} G_{\vec{j}}:= G_{j_1} \otimes \cdots \otimes
G_{j_4} : \scrT^{(\vec{j},\vec{j})} \to \scrT^{(\vec{j},\vec{j})}
\end{equation}
where $\scrT^{(\vec{j},\vec{j})}$ is the tensor space introduced in
(\ref{tensp_def}).  One can see immediately from its definition that
\begin{equation}
\label{nodeG_im} \mIm G_{\vec{j}} \subseteq \Hil^{FKLS, \; GF}_{n}
\end{equation}
But from equation (81) in \cite{fk2007},
\begin{equation}
\label{reduced_res} G_j = d_j^2 \sum_{k=0}^{2j} C^j_k P_k
\end{equation}
where $P_k : \Hil_{(j,j)} \to \Hil_{(j,j)}$ is the projector onto
the spin $k$ representation in the decomposition of $\Hil_{(j,j)}$
into irreducibles of the fixed $SO(3)$ subgroup ($\Hil_{(j,j)} =
\oplus_{k=0}^{2j} \Hil_{k}$), and where $C^j_k$ is given by
\begin{equation}
\label{FKLScoef}
C^j_k = \frac{(2j)!}{(2j-k)!}\frac{(2j)!}{(2j+k+1)!} .
\end{equation}
In (\ref{reduced_res}), because all of the coefficients of the
projection operators are non-zero, $G_j$ is manifestly invertible
and hence its image is the entirety of $\Hil_{(j,j)}$.  It follows
that the image of $G_{\vec{j}}$ is the entirety of
$\scrT^{(\vec{j},\vec{j})}$.  Thus, from (\ref{nodeG_im}), the gauge
fixed solution space for FKLS at the node $n$ is the entirety of
$\scrT^{(\vec{j},\vec{j})}$. From this, in turn, it follows that,
after projecting onto the $SO(4)$ gauge-invariant subspace, we will
obtain
\begin{equation}
\Hil_{n}^{FKLS} = \scrI^{(\vec{j},\vec{j})},
\end{equation}
i.e., the final solution space for the intertwiners in the FKLS
model is in fact \textit{all of the $SO(4)$ intertwiners}.

Thus, if one is to understand the implementation of cross-simplicity
in this model as a restriction on states (in particular, boundary
states), then in fact we see that \textit{no constraint is imposed}.
The first significance of this is that it shows clearly that the
boundary Hilbert space of the FKLS model is not isomorphic to the
Hilbert space of LQG, whereas there are strong indications that the
boundary Hilbert space of the flipped model \cite{eprlett, eprpap}
is so isomorphic.  Secondly, it raises questions as to the method
used to incorporate cross-simplicity in FKLS.  It is true there is
still a sense in which cross-simplicity is imposed: cross-simplicity
still affects the final vertex amplitude in the FKLS model. However,
it is unusual for a constraint to only manifest itself in this way.

One more remark is in order.  As noted in subsection \ref{newmodels},
the expectation values of the constraint operators $\hat{C}^j_f$ with
respect to the FKLS coherent states \Ref{fkls_int_gf} are in fact zero.
However, the
\textit{matrix elements} of these constraint operators with
respect to these states are \textit{not} zero.  This highlights
the importance of satisfying the constraints by matrix elements, and
not just by expectation values: for the former is closed under
linear combinations, whereas the latter is not. The mere fact that
the constraints $\hat{C}^j_f$ have zero expectation value on the ``solution''
states \Ref{fkls_int_gf} tells us nothing about their span, and in fact their
span is the whole intertwiner space.

\section{The coherent state approach and first class constraints}

We propose that the reason the coherent state approach
fails to constrain the states in FKLS is due to the first class
nature of the cross-simplicity system of constraints in this case.
More generally, we suggest that one should be careful when using the 
coherent state 
method when the constraints are not fully second class, as it may 
underimpose the constraints on the state space. We support such 
a claim with a simple example that we hope
captures the essence of the problem.

\subsection{A simple example}

A simple example suffices to show why this should be the
case.  In order to keep things conceptually clear,
we first consider a single simple harmonic oscillator with phase
space $\{(q,p)\}$ and the standard Poisson brackets
$\{q,p\}=1$.  The kinematical quantum state space is then the standard
$\Hil_{kin} = L^2(\R)$, with $\hat{q}$ and $\hat{p}$ acting
in the usual way by multiplication and derivation. Consider the standard
(unnormalized) family of coherent
states for the simple harmonic oscillator:
\begin{equation}
\label{cohstatedef}
\psi^{coh}_{(q_o,p_o)}(q) = e^{-ip_o q} e^{-(q-q_o)^2/2}.
\end{equation}
Consider the pair of second class constraints
\begin{equation}
\label{secclass}
q=0 \qquad \text{and} \qquad p=0 .
\end{equation}
The solution space for this pair of constraints, using
the coherent states \Ref{cohstatedef} is
\begin{equation}
\Hil^{\text{2nd class}}_{phys} := {\rm span}\{\Psi^{coh}_{(0,0)}\}
\end{equation}
i.e., the one dimensional space spanned by the vacuum.
Thus, one obtains an actual constraint on the state space,
and it is in fact what one expects: classically the
constraints \Ref{secclass} completely constrain the
phase space to a point.  Likewise, in the final quantum
theory, there is only a single state, modulo rescaling.
Consider now instead the first class system
\begin{equation}
\label{firstclass}
q=0 .
\end{equation}
The solution space of this constraint, using
the coherent states \Ref{cohstatedef} is
\begin{equation}
\Hil^{\text{1st class}}_{phys} :=
\overline{{\rm span}\{\Psi^{coh}_{(0,p_o)}\}_{p_o\in \R}}
= \overline{{\rm span}\{e^{-ip_o q} e^{-q^2/2}\}_{p_o \in \R}}.
\end{equation}
where the overbar denotes Cauchy completion.
It is not hard to see that in fact
$\Hil^{\text{1st class}}_{phys} = \Hil_{kin}$.
\begin{quote}
\textit{Proof}: We show that the orthogonal complement of
$\Hil^{\text{1st class}}_{phys}$ in $\Hil_{kin}$
is trivial.
Suppose $\Psi$ is orthogonal to $\Hil^{\text{1st class}}$.
Then $\int \dif q e^{ip_o q} e^{-q^2/2} \Psi(q) = 0$ for all $p_0$.
Then the Fourier transform of $e^{-q^2/2}\Psi(q)$ is zero, so
that $\Psi(q)$ is zero. $\Box$
\end{quote}
Thus no constraint has been imposed in $\Hil^{\text{1st class}}_{phys}$.

Now, one may think at first glance that this is a rather trivial
example.  However, one can easily extend it to an arbitrary number
$n$ of simple harmonic oscillators.  In this case the kinematical
Hilbert space $\Hil_{kin}$ decomposes into a tensor product of
kinematical Hilbert spaces, one for each oscillator:
\begin{equation}
\Hil_{kin} = \otimes_{i=1}^n \Hil^i_{kin}
\end{equation}
and the coherent states $\Psi^{coh}_{(\vec{q}_o, \vec{p}_o)}$
correspondingly decompose into a tensor product of coherent states
\begin{equation}
\label{multicoh}
\Psi^{coh}_{(\vec{q}_o, \vec{p}_o)}
= \otimes_{i=1}^n \Psi^{coh}_{(q_o^i,p_o^i)} .
\end{equation}
Finally, for some $m < n$, one can consider
either the second class set of constraints
\begin{equation}
\label{multisecond}
q_i = 0 \qquad \text{and} \qquad p_i=0 \qquad \text{for }i=1,\dots,m
\end{equation}
or the set of first class constraints
\begin{equation}
\label{multifirst}
q_i = 0 \qquad \text{for }i=1,\dots,m .
\end{equation}
Imposing these two sets of constraints using the
coherent states \Ref{multicoh} then yields conclusions similar
to those for the single simple harmonic oscillator.
For the case of the second class constraints \Ref{multisecond},
the state space is reduced to that of $n-m$ simple
harmonic oscillators, as one would expect.  However,
for the case of the first class constraints \Ref{multifirst},
no constraint is imposed on the state space.

\subsection{The classes of constraints in the flipped case}

One can also repeat the multiple-simple-harmonic-oscillator example
of the last subsection, but mix the first and second class
constraints.  If one then imposes the constraints using the coherent
states, one finds that the state space is reduced, but not as much
as it should be.  This then raises a question. In the case of the
\textit{flipped} model, it is true that the cross-simplicity
constraints are not first class. However, as shown in appendix
\ref{alg_app}, neither are they \textit{purely} second class. There is a
first class component given by
\begin{equation}
\label{firstcl_comp} \tilde{C}_f \propto L_f^i C_{fi} = L_f^i K_{fi}
= C_{ff}
\end{equation}
But as one can see, this first class component is just the diagonal
simplicity constraint for the link.  The coherent states
\Ref{fl_int_gf} are sharply peaked with respect to this constraint,
and so use of the coherent state method does not cause ``spread'' in
this constraint\footnote{This is partially related to the fact that
the coherent states \Ref{fl_int_gf} (and \Ref{fkls_int_gf}) are only
partially coherent states: they are coherent states with respect to
the operators $\hat{B}_f(t)^{IJ}$, but not with respect to the
connection operators $\hat{U}_f$.  This is why it is possible for
the coherent states to be sharply peaked with respect to $\hat{B}
\cdot \hat{B} = \hat{C}^f_4$ and $\dual\hat{B}\cdot\hat{B} =
\hat{\tilde{C}}^f_4$.}. As a consequence, the coherent state method,
in this case, actually imposes \Ref{firstcl_comp} strongly, as it
should\footnote{Showing how extraordinarily well-adapted the coherent
states \Ref{fl_int} are for imposing these constraints.}.

The master constraint method is ``smart enough'' to impose first
class constraints strongly and second class constraints weakly, so
that the existence of a first class component is of no concern for
the master constraint method. The fact that the coherent state
method and master constraint method match for the flipped model can
thus be taken as a further check that the chosen coherent states
\Ref{fl_int_gf} handle correctly the issue of the class of the
constraints.

\section{The spectra of the area operators}

In this section, we will analyze the spectra of area operators in
the two models, and compare these spectra with those found in loop
quantum gravity.  We will find that the area spectrum of the model
for small $\gamma$ \cite{eprlett, eprpap, ls_model} matches that of
loop quantum gravity.  The area spectrum of the model for infinite
$\gamma$ \cite{fk2007, ls_model} (FKLS model) will be found to be
quite different.

The section is organized as follows. First, a derivation of the
spectra of the area operators for general $\gamma$ will be given.
For $\gamma \ll 1$ and $\gamma = \infty$, we then show that the LQG
spectrum, or, respectively, that of \cite{sergei2007} is obtained.

\subsection{Areas: Classical analysis}

Classically, the area of a triangle $f$ is given by
\footnote{To see this, choose Cartesian coordinates $(x^1,x^2)$
in the triangle $f$; then
\begin{displaymath}
(\dual B_f) ^{IJ} (\dual B)_{fIJ} =\frac{1}{2}\left(e_1\cdot e_1\;
e_2\cdot e_2-\left(e_1\cdot e_2\right)^2\right)=2\;Area(f)
\end{displaymath}}
\begin{equation}
\label{SO4area} A_4(f)^2 = \frac{1}{2} (\dual B_f)^{IJ} (\dual
B_f)_{IJ} = \frac{1}{2} B_f^{IJ} B_{fIJ} =
\frac{2{\Gfactor}^2\gamma^4}{\left(\gamma^2-1\right)^2}\;\left[
\left(J_f^{(+)i}\right)^2\left(1-\frac{1}{\gamma}\right)^2+
\left(J_f^{(-)i}\right)^2\left(1+\frac{1}{\gamma}\right)^2\right]
\end{equation}
When the gauge-fixed cross-simplicity constraints \Ref{GF_3ii}
hold, the $B_f^{ij}$ vanish and the above quantity is equal to
\begin{equation}
\label{SO3area} A_3(f)^2 := B_f^{0i} B_{f0i} = {\Gfactor}^2
\frac{\gamma^4}{\left(\gamma^2-1\right)^2}\;
\left[J_f^{(+)i}\left(1-\frac{1}{\gamma}\right)
-J_f^{(-)i}\left(1+\frac{1}{\gamma}\right)\right]^2
\end{equation}
which we refer to as the gauge-fixed area of $f$.
Rewriting (\ref{SO3area}) in preparation for
quantum theory, using $L^i = J^{(+)i}+J^{(-)i}$,
we have
\begin{equation}
A_3(f)^2 = 2{\Gfactor}^2 \frac{\gamma^4}{(\gamma^2-1)^2}\left[(J^{(+)i})^2
\left(1-\frac{1}{\gamma}\right)
+(J^{(-)i})^2\left(1+\frac{1}{\gamma}\right)\right] - k^2
\;\frac{\gamma^2}{\gamma^2-1}(L^i)^2
\end{equation}

\subsection{Areas: Quantum analysis}

Let us take a look at the quantum area operators. Working in the
$SO(4)$ (generalized) spin-network basis,
$\hat{A}_4(f)$ is diagonal with eigenvalues
\begin{equation}
spec(\hat{A}_4(f)^2)=\left\{2{\Gfactor}^2\left[\frac{\gamma^2}{(\gamma+1)^2}\;j^+(j^+
+1)+\frac{\gamma^2}{(\gamma-1)^2}\;j^-(j^- +1)\right]\;\; |\;\;
j^\pm \in \mathbb{N}/2 \right\}
\end{equation}
The spectrum of $\hat{A}_3(f)$
can be easily calculated and is given by:
\begin{eqnarray}
\nonumber
spec(\hat{A}_3(f)^2)&=&\left\{\;\frac{2{\Gfactor}^2\gamma^4}{(\gamma^2-1)^2}\left[j^+(j^+
+1)\left(1-\frac{1}{\gamma}\right)+j^-(j^-
+1)\left(1+\frac{1}{\gamma}\right)\right]\right. \\
&& \qquad \qquad \qquad \qquad \qquad \qquad
-\left.\;\frac{{\Gfactor}^2\gamma^2}{\gamma^2-1}\;k(k+1)\;\; |\;\;
j^\pm \in \mathbb{N}/2\; ,\; k\in \mathbb{N}\;\right\}
\end{eqnarray}
For small $\gamma$ and infinite $\gamma$, diagonal simplicity dictates $j_+ =
j_- \equiv j$($\in \N/2$). In these cases, the spectrum of
$\hat{A}_4(f)^2$ reduces to $4{\Gfactor}^2\gamma^2 j(j+1)$ for
$\gamma \ll 1$, and to $4{\Gfactor}^2j(j+1)$ for $\gamma = \infty$. The
spectrum of $\hat{A}_3(f)^2$ reduces to
\begin{eqnarray}
\label{smallgspec}
\gamma \ll 1 \quad &\rightarrow& \quad spec(\hat{A}_3(f)^2)=\left\{{\Gfactor}^2\gamma^2\;k(k+1)\;\; | \;\; k\in \mathbb{N}\right\} \\
\label{biggspec} \gamma = \infty \quad &\rightarrow& \quad
spec(\hat{A}_3(f)^2)=\left\{4{\Gfactor}^2 \; j(j+1)-{\Gfactor}^2 \;
k(k+1) \;\; | \;\; j=j^+=j^- \in \mathbb{N}/2\; ,\; k\in
\mathbb{N}\right\}
\end{eqnarray}
%
%
It is interesting to note that in the small $\gamma$ case, the
dependence of the spectrum on $\gamma$ is exactly the same as in
LQG, whereas in the large $\gamma$ limit, all dependence of the
spectrum on $\gamma$ vanishes. Furthermore, note the latter spectrum
is exactly the Euclidean analogue of the Lorentzian area spectrum
presented in equation (24) of \cite{sergei2007}.  This is in part
not surprising, as, in \cite{sergei2007}, the $B$ variables are
quantized as the generators of $SO(4)$, which, as can be seen from
\Ref{Bquant}, is the symplectic structure corresponding to the
$\gamma = \infty$ case\footnote{Furthermore, in (24) in
\cite{sergei2007}, one has not yet fully imposed cross-simplicity,
as in equation \Ref{biggspec} here one has not yet imposed
cross-simplicity.}.
%
%

Incorporation of the cross-simplicity constraints in the BC model
leads to the area spectrum \Ref{biggspec} with $k$ set to zero.
Incorporation of the cross-simplicity constraints for the flipped
and FKLS models, however, does not change the spectra \Ref{smallgspec},
\Ref{biggspec}. In the flipped case, cross-simplicity tells us that $k=2j$,
whereas in the FKLS model, as discussed in the last two sections,
there are no restrictions on the space of states from
cross-simplicity.

Let us next recall that in the classical theory, the gauge-fixed
area $A_3(f)$ equals the non-gauge-fixed area $A_4(f)$. What happens
to this equality in quantum theory?  The fate of this equality is
different in the FKLS, flipped, and BC models; we first state what
happens in each of the models and then shed light on why. First, in
the FKLS model, the spectra are completely different. In the flipped
model, the spectra differ only by an order of $j$, a term which is
zero in the semiclassical limit.  Finally, in the BC model, the
spectra are exactly equal even in the quantum theory.  To see why
these observations are true, we first note that the difference
between the gauge-fixed and non-gauge-fixed areas is proportional to
the sum of the squares of the gauge-fixed cross-simplicity
constraints \Ref{GF_3ii} --- the \textit{master constraint} $M_{tf}$ for
cross-simplicity at a given $t$ and $f$.  It is then easy to see why
in the FKLS case the spectra are completely different: in FKLS,
cross-simplicity is not imposed at all as a restriction on states,
so that there is no restriction on the eigenvalues of
$\hat{M}_{tf}$. In the flipped model, one does impose
cross-simplicity on states, and as noted earlier, the method of
imposing cross-simplicity can even be viewed as choosing the minimal
eigenvalue of the master constraint.  The minimal eigenvalue is a
quantity zero in the semiclassical limit, and this is why, in the
flipped model, the spectra of the gauge-fixed and non-gauge-fixed
areas differ by a quantity that is zero in the semiclassical limit.
Finally, in the BC model, the cross-simplicity constraints \Ref{bigg_3ii}
are imposed strongly, directly as operator equations, so that the difference
between the operators $\hat{A}_4(f)$ and $\hat{A}_3(f)$ is exactly
zero after imposing of cross-simplicity.

It remains to address a natural question. $\hat{A}_3(f)$ is not an
$SO(4)$-Gauss gauge invariant quantity. One may then ask: why are we
then interested in the the spectra (\ref{smallgspec}),
(\ref{biggspec})? As noted in footnote \ref{timegaugefn} above, the
gauge-fixing involved in defining $A_3(f)$ is a part of the time
gauge which is used in LQG. Thus it is natural to look at the area
$A_3(f)$ when comparing spectra with those in LQG. Furthermore, we
are considering these spectra in the spirit of \cite{carlospec}.
That is, $\hat{A}_3(f)$ is viewed as a partial observable: it is to
be made into a complete observable by coupling it with an
appropriate choice of clock \cite{carlospec, partialobs} (and
\textit{not} by group averaging it).
%
%
In the present case, the relevant gauge freedom is the internal tetrad
degrees of freedom; therefore, an appropriate clock would have to be
constructed from a field that is sensitive to this internal freedom,
such as a spinorial matter field.
The spectrum of the resulting complete observable will probably depend
on which clock is used.
The kinematical spectrum is then seen as a ``clock-independent
spectrum''; its physical meaning is discussed, e.g., in
\cite{carlospec}.

\section{Discussion}

In this paper, we have discussed properties of the two recently
proposed spin-foam models \cite{eprlett, eprpap} and
\cite{fk2007,ls_model}, which we refer to as the flipped model and
FKLS model, respectively. In particular we have recalled that the
boundary Hilbert space of the flipped model matches the Hilbert
space of LQG, whereas we have shown that the boundary space of the
FKLS model is completely different. The area spectra of the two
theories were also compared, and it was found that in the flipped
case, the ($SO(3)$) area spectrum exactly matches that of LQG,
whereas in the FKLS model, the spectra are completely different.  In
the case of the BC model, the boundary Hilbert space differs from
that of LQG in that the BC model has no intertwiner degrees of
freedom; and the BC area spectrum differs from that of LQG --- it 
does not depend at all on the Barbero-Immirzi parameter.

Furthermore, an unusual aspect of the FKLS model was pointed out: in
the FKLS model, cross-simplicity is in fact not imposed on states.
This was seen to be due to the fact that the gauge-fixed
cross-simplicity constraints in fact become a first class system,
and FKLS uses a coherent state method in imposing them.  That is to
say, not only does the first class nature of the constraints tell us
that the use of coherent states is not needed (since the constraints
can now be imposed strongly), it also tells us that the use of
coherent states will lead to no constraint being imposed on states.
Nevertheless, it is possible that this is not a fatal problem. For,
the strategy used in \cite{fk2007, ls_model} to construct the FKLS
model is such that cross-simplicity has an effect on the final
vertex of the model. More specifically, the vertex seems to suppress
intertwiners that are far from the BC intertwiner. This heuristic
statement comes from the fact that the coefficients $C^j_k$ in
\Ref{FKLScoef} appear in the FKLS amplitude sum:
\begin{equation}
\sum_{j_{f}} \prod_{f}{\mathrm{d}_{j_{f}}^{2}}
\sum_{l_{t}}\prod_{t}\mathrm{d}_{l_{t}}
\sum_{k_{tf}}\prod_{tf}{\mathrm{d}_{j_{f}}^{2}}C^{j_{f}}_{k_{tf}}
A^{Grav}(j_{f},k_{tf}, l_{t})
\end{equation}
where $A^{Grav}(j_{f},k_{tf}, l_{t})$ is as in equation
(84) of \cite{fk2007}.
The coefficients $C^j_k$ are peaked for $k=0$ and vanish for large $k$.
$k=0$ corresponds to the Barrett-Crane intertwiner, whence it appears that
the FKLS model favors propagation of the Barrett-Crane intertwiner.
This seems to be the sense in which FKLS imposes cross-simplicity:
in this light, one can see how FKLS is a weakened version of the
Barrett-Crane model, but not weakened to the point of being BF theory.
Nevertheless, as we have said, this way of imposing constraints is
somewhat unusual.

In summary, the flipped model exactly matches LQG at the level of
Hilbert space structure and area operators, whereas the FKLS model
and BC models do not.  Furthermore, in the FKLS model, it was found
that no cross-simplicity is imposed on states, raising questions as
to how much its quantization of these constraints should be trusted.
Forthcoming papers are in preparation on the Lorentzian case and the
case of arbitrary $\gamma$ \cite{flippedlor,gengamma}; these will
further clarify these issues.

%
%

\section*{Acknowledgements}
We thank Carlo Rovelli, Alejandro Perez, Simone Speziale, Etera
Livine, and Laurent Freidel for valuable exchanges.  We also thank
Jorge Pullin for inviting one of us (JE) to give an International
Loop Quantum Gravity Seminar on the flipped vertex; this paper arose
in part from the desire to clarify some of the issues raised in that
seminar. JE was supported by an NSF International Research
Fellowship under grant OISE-0601844.

\appendix

\section{Poisson algebra of the off-diagonal simplicity constraints}
\label{alg_app}

In this appendix we first recall some facts about the structure of
the algebra $\so(4)$ and then analyze the class structure of the
constraints \Ref{GF_3ii}.

Let $J^{IJ}$ be the generators of $\lalg{so(4)}$. Concretely, one
can take $\so(4)$ to be the matrix algebra of skew 4 by 4 matrices;
in terms of this viewpoint, the generators $J^{IJ}$ are defined to
be the basis $\left(J^{IJ}\right)^M{}_N := 2
\delta^{M[I}\delta^{J]}{}_N$, with the Lie bracket given by the
matrix commutator.

Define the selfdual and anti-selfdual generators $J_\pm := *J \pm J,
$ that satisfy $J_\pm =\pm  *J_\pm$. Then it is immediate to see
that $[J_+,J_- ]=0$. The $J_+$ span a three dimensional subalgebra
$\lalg{su(2)}_+$ of $\lalg{so(4)}$, and the $J_-$ span a three
dimensional subalgebra $\lalg{su(2)}_-$ of $\lalg{so(4)}$, both
isomorphic to $\lalg{su(2)}$. It is convenient to choose a basis in
$\lalg{su(2)}_+$ and in $\lalg{su(2)}_-$.   For this, choose a unit
norm vector $n$ in $R^4$, and three other vectors $v_i, i=1,2,3$
forming, together with $n$, an orthonormal basis, and define
\begin{equation}
J_\pm^i = \half({}^*J\pm J)_{IJ}\ v_i^I n^J
\end{equation}
The $\lalg{su(2)}$ structure is then easy to see, since
$[J_\pm^i,J_\pm^j]= \epsilon^i{}_{jk} J_\pm^{k} $. In particular, we
can choose $n=(0,0,0,1)$, and $v_i^I=\delta^I_i$, and we have
\begin{equation}
J_\pm^i = \frac{1}{4} \epsilon^i{}_{jk}J^{jk} \pm \half J^{0i} .
\label{basis}
\end{equation}
The Casimirs $(J^{(+)i})^2$, $(J^{(-)i})^2$ have spectra $\{j^+(j^+
+ 1) \mid j^+ \in \N/2\}$ and $\{j^-(j^- + 1) \mid j^- \in \N/2\}$,
respectively, and irreducible representations of $\SO(4)$ are
labeled by pairs $(j^+,j^-)$.

Finally, the generators of rotations and boosts in the chosen frame
$n^I = \delta^I_0$, $v_i^I = \delta^I_i$ are
\begin{eqnarray}
L^i &=& \frac{1}{2}\epsilon^i{}_{jk}J^{jk} = J^{(+)i} + J^{(-)i} \\
K^i &=& J^{0i} = J^{(+)i}-J^{(-)i}
\end{eqnarray}
and they satisfy the algebra
%
%
\begin{eqnarray}
\left[ L^i , L^j \right] &=& \epsilon^{ij}{}_k L^k \\
\left[ K^i , K^j \right] &=& \epsilon^{ij}{}_k L^k \\
\left[ L^i , K^j \right] &=& \epsilon^{ij}{}_k K^k
\end{eqnarray}
With these formulae, it is easy to analyze the algebra of the gauge
fixed off-diagonal simplicity constraints \Ref{GF_3ii},
$C_f^i=L_f^i-\frac{1}{\gamma}K_f^i$.
\begin{equation}
\label{pbmatrix}
\{C_f^i, C_f^j\} =
\epsilon^{ij}{}_k\left\{\left(1+\frac{1}{\gamma^2}\right)L^k_f -
\frac{2}{\gamma}K^k_f \right\} .
\end{equation}
The first thing we note from this expression is that, for finite
non-zero $\gamma$, the algebra of constraints $\{C_f^i\}$ does not
close and so is not first class.  For $\gamma \ll 1$, the
constraints $C_f^i$ are proportional to the boost generators $K_f^i$
and so the constraint algebra again fails to close.  For $\gamma = \infty$,
%
%
on the other hand, $C_f^i = L_f^i$, so that the algebra closes
and forms a first class system.

Let us investigate further the cases of finite $\gamma$ and of $\gamma \ll 1$.
In these cases the constraints are second class, but we know that
they cannot be \textit{purely} second class, as the constraints $C_f^i$
are three in number, and second class constraints always come in pairs.
Thus, there must exist a first class component in the constraints.
In fact, one can determine the first class component from \Ref{pbmatrix}.
The (phase space dependent) null vector of the
poisson bracket \Ref{pbmatrix} matrix can be read off as
\begin{equation}
X_f^i = \left(1+\frac{1}{\gamma^2}\right)L^k_f -
\frac{2}{\gamma}K^k_f
\end{equation}
so that the first class component of the cross-simplicity
constraints is
\begin{eqnarray}
\tilde{C}_f &=& X_f^i C_{fi} =
\left(1-\frac{1}{\gamma^2}\right)L_f^2 +
\frac{2}{\gamma}\left(K_f^2+L_f^2\right) -
\frac{1}{\gamma}\left(3+\frac{1}{\gamma^2}\right)K_f \cdot L_f
\end{eqnarray}
For small $\gamma$ this reduces to
\begin{equation}
\tilde{C}_f \propto L_f^i C_{fi} \propto L_f \cdot K_f ,
\end{equation}
which is just the diagonal simplicity constraint for small $\gamma$.

\section{Coherent states}
\label{app_coh}

In this appendix we review the construction of coherent states for
$SU(2)$ and $SO(4)$ and derive the coefficients $c_i(n_a)$ referred
to in the main text. For details see \cite{ls_cohstates,perelomov}.
Coherent states for $SU(2)$ are defined as follows. Consider the
carrying space $\mathcal{H}_j$ of the spin $j$ representation of
$SU(2)$. This is spanned by the states $\left\{|j,m\rangle\right\},
m\in \mathbb{N}/2,|m|\leq j$. Consider now the states invariant
under the $U(1)$ subgroup that generates rotations around the $z$
axis $\hat{e}_z$. Any element of the basis above satisfies this
requirement. Next, we select from among these states those that
minimize the uncertainty
$\Delta^2:=\langle\vec{J}^2\rangle-\langle\vec{J}\rangle\cdot\langle\vec{J}\rangle$.
There are two such states: $|j,\pm j\rangle$. Coherent states are
defined by acting with $SU(2)/U(1)\sim S^2$ on each of these states.
Of the two states $|j,\pm j \rangle$, it is thus sufficient to
restrict consideration to $|j,j\rangle$, as the other can be
obtained by a rotation of angle $\pi$ around any vector in the $xy$
plane. For each unit vector $\hat{n}\in S^2$, we then define the
coherent state \be \label{cohdef} |j, \hat{n}\rangle:=g(\hat{n})|j,
j\rangle \ee where $g(\hat{n})\in SU(2)$ denotes the group element
that rotates $\hat{e}_z$ into the direction $\hat{n}$. Explicitly,
if, in coordinates, $\hat{n}$ is given by $\hat{n}=\left(\sin\theta
\cos\phi,\sin\theta\sin\phi,\cos\theta\right)$, then $g(\hat{n})$
may be taken to be $\exp(i\theta\hat{m}\cdot\vec{J})$, where
$\hat{m}:=\left(\sin\phi ,-\cos\phi ,0\right)$ is a unit vector
orthogonal to both $\hat{e}_z$ and $\hat{n}$, and $\vec{J}$ are the
generators of the algebra.  These coherent states so defined clearly
form an overcomplete basis of the space $\mathcal{H}_j$.

The decomposition of the coherent state $|j, \hat{n}\rangle$ in the
usual basis $|j,m\rangle$ is given by: \be
|j,\hat{n}\rangle=g(\hat{n})|j, j\rangle=\sum_m\;
D^j_{m,j}(g(\hat{n}))|j,m\rangle \ee where $D^j(g)$ denotes the
representation matrix of the group element $g$ acting on the
carrying space $\mathcal{H}_j$. Furthermore, the following identity
will be useful in what follows: \be |j,-\hat{n}\rangle=\sum_m\;
\bar{D}^j_{mj}(g(\hat{n}))(-1)^{j+m}|j,-m\rangle \label{conjugate}
\ee where we have used the fact that
$\bar{D}^j_{mn}(g)=(-1)^{m+n}D^j_{-m-n}(g)$. We also give some
useful formulas for calculating expectation values:
\begin{equation}
\langle j,\hat{n}|J^i|j,\hat{n}\rangle=jn^i;
\end{equation}
and
\begin{equation}
\langle j, \hat{n}|
(J^i)^2|j,\hat{n}\rangle=\frac{j}{2}+j\left(j-\frac{1}{2}\right)(n^i)^2,
\quad \text{for} \; i=x,y,z.
\end{equation}
In particular this allows to
compute: \be
\Delta^2=\langle\vec{J}^2\rangle-\langle\vec{J}\rangle\cdot\langle\vec{J}\rangle=j.
\ee Now, using the local isomorphism $SO(4)\sim_{loc.} SU(2)\times
SU(2)$, one can use $SU(2)$ coherent states to define $SO(4)$
coherent states. We then define $SO(4)$ coherent states as given by
the action of $SU(2)\times SU(2)/U(1)\times U(1)\sim S^2\times S^2$
on the states $|j^+,j^+\rangle\otimes|j^-,j^-\rangle$:
$|j^+,\hat{n}^+\rangle\otimes|j^-,\hat{n}^-\rangle
:=(g(\hat{n}^+),g(\hat{n}^-))|j^+,j^+\rangle\otimes|j^-,j^-\rangle$.
These states form an overcomplete basis of the carrying space
$\mathcal{H}_{j^+}\otimes\mathcal{H}_{j^-}$. Among these there are
two classes of states that will be of more interest, given by
$j^+=j^-$ and $\hat{n}^+=\pm\hat{n}^-$, as they solve the simplicity
constraints for the case of flipped and unflipped symplectic
structure, as noted in the main text.

We are now ready to calculate the coefficients $c_i(\hat{n}_a)$
given in the main text. Remember the form of the intertwiner for
both the flipped ($I^+$) and the FKLS ($I^-$) models:
\begin{equation}
I^\pm=\int_{SO(4)}dG\;
G\cdot\bigotimes_a\left(|j_a,\hat{n}_a\rangle\otimes|j_a,\pm\hat{n}_a\rangle\right).
\end{equation}
Let us first write these states in the $|j,m\rangle$ basis as above:
\be I^\pm=\int_{SU(2)\times SU(2)}dg^+dg^-\;
\bigotimes_a\left(D^{j_a}_{m_a^+ n_a^+}(g^+)D^{j_a}_{n_a^+ j}(g_a)\;
D^{j_a}_{m_a^- n_a^-}(g^-)D^{j_a}_{n_a^- \pm
j}(g_a)\;|j_a,m_a^+\rangle\otimes|j_a,m_a^-\rangle\right) \ee where
we have written $G=(g^+,g^-)$, $g_a:=g(\hat{n}_a)$ and summation
over repeated indices is understood. Integration over $g^\pm$ gives
four valent intertwiners of $SU(2)$, $C^{i^\pm}_{m_1^\pm ...
m_4^\pm}$. Defining $|i^\pm
;j_a\rangle:=C^{i^\pm}_{\{m_a^\pm\}}\bigotimes_a|j_a,m_a^\pm\rangle$,
one can rewrite the last equation as: \be
I^\pm=\sum_{i^+,i^-}\left(C^{i^+}_{\{n_a^+\}}\bigotimes_a
D^{j_a}_{n_a^+
j}(g(\hat{n}_a))\right)\left(C^{i^-}_{\{n_a^-\}}\bigotimes_a
D^{j_a}_{n_a^- \pm j}(g(\hat{n}_a))\right)\;\;
|i^+;j_a\rangle\otimes|i^-;j_a\rangle. \ee One can then readly read
the form of the coefficients $c_{i^+}(\hat{n}_a)$: \be
c_{i^+}(\hat{n}_a)=C^{i^+}_{\{n_a^+\}}\bigotimes_a D^{j_a}_{n_a^+
j}(g_a). \ee The fact that for $I^-$ the coefficients for the anti
self dual part are conjugated in eq. \Ref{unfl_int_sdasd} comes from
eq. \Ref{conjugate} above and some elementary properties of the
Clebsch-Gordan coefficients.

\section{Fixing the scaling of the symplectic structure}
\label{scale_app}

In this appendix we fix the scaling of the symplectic structure
in the classical discrete theory and so determine the
correct coefficient in \Ref{j_interms_b}.

There are two possible ways to fix the coefficient in the symplectic
structure: (1.) by first fixing the coefficient in the discrete
action and then deriving the symplectic structure from the discrete
action, or (2.) by setting the Poisson brackets of the discrete
variables equal to their Poisson brackets in the corresponding
continuum theory.  The former, although seemingly more systematic,
is problematic, first because the manner of fixing the coefficient
in the discrete action is not fully understood\footnote{Naively one
would think one could fix the scaling in the action by requiring the
discrete action to be approximately equal to the corresponding
continuum action.  However, there is a difficult with this. In the
derivation in \cite{eprpap}, if one is more careful with numerical
factors, one finds that in relating the bulk discrete action to the
bulk continuum action, there is a missing factor of 12, because the
shape of the elementary volume associated to each face in the sum
was not taken into account.  However, in comparing the discrete
boundary action with the continuum boundary action, there was a
missing factor of 3 (again because of the shape of the elementary
volumes in the sum). But if one corrects the discrete bulk and
boundary actions with these factors, so that they equal their
continuum counterparts in the continuum limit, then the action will
no longer be approximately additive, a requirement noted in
\cite{eprpap} and \cite{hartlesorkin}.}, and second because the
manner of deriving the symplectic structure from the action in
\cite{eprpap} is not completely standard.


Thus, we carry out the second option. The symplectic structure of
the discrete theory is implicitly specified by equation
\Ref{j_interms_b} in the main text, $J_f(t)^{IJ}$ being defined to
be the phase space function whose hamiltonian vector field is the
left-invariant vector field, on the group $U_f(t',t)$, corresponding
to the Lie algebra element $J^{IJ}$ defined in appendix
\ref{alg_app}. We begin by replacing equation \Ref{j_interms_b} with
\begin{equation}
\label{scaled_jb}
J_f(t) = \lambda\left(B_f(t) + \frac{1}{\gamma}\dual B_f(t)\right)
\end{equation}
where now $\lambda \in \R$ is a coefficient to be fixed by
comparison with the continuum theory (onshell with respect to
the Gauss constraint/closure constraint). We will show
that this manner of fixing the coefficient results
in precisely the coefficient appearing in \Ref{j_interms_b} in
the main text.

In the continuum BF theory, we start from the action
\begin{equation}
\label{cont_action}
S = \frac{1}{2\Gfactor} \int_{\scrM} \left[B_{IJ} \wedge F^{IJ} +
\frac{1}{\gamma} (\dual B)_{IJ} \wedge F^{IJ} \right]
\end{equation}
where $B^{IJ}_{ab} = B_{ab}^{[IJ]}$ is a two-form and
$F^{IJ}_{ab}$ is the curvature of an
$SO(4)$ connection $\omega^{IJ}_a$.
%
%
Upon substituting in $B^{IJ}= \dual e^I \wedge e^J$, this becomes
the Holst action with all of the correct numerical factors \cite{alrev}.
To simplify the following derivations from the action, define
\begin{equation}
P^{IJ}:=
\frac{1}{2\Gfactor}\left(B^{IJ}+\frac{1}{\gamma}(\dual B)^{IJ}\right)
\end{equation}
Following the prescription for deriving
symplectic structure used in \cite{alrev} and \cite{asht_et},
we vary the action \Ref{cont_action} to obtain the symplectic one-form
\begin{equation}
\Theta(\delta) = \int_{\Sigma} P_{IJ} \wedge \delta \omega^{IJ}
\end{equation}
The symplectic structure of the theory is then
\begin{equation}
\Omega(\delta_1, \delta_2) =
\int_{\Sigma} \left[\delta_1 P_{IJ} \wedge \delta_2 \omega^{IJ}
- (1 \leftrightarrow 2) \right]
\end{equation}
so that the basic non-trivial Poisson brackets in the continuum theory
are
\begin{equation}
\{A_a^{IJ}(\vec{x}), P_{bcKL}(\vec{y})\} = \epsilon_{abc}
\delta^{[I}_K \delta^{J]}_L \delta^3(\vec{x},\vec{y})
\end{equation}
where $\epsilon_{abc}$ is the Levi-Civita symbol (Levi-Civita tensor
density of weight $-1$). Given an edge $\ell$, with source point
$n$, let $U_{\ell}(n)$ denote the parallel transport defined by
$\omega_a^{IJ}$ along $\ell$, starting at $n$. Given an oriented
2-surface $S$, define the ``flux'' $P(S)^{IJ}$ by
\begin{equation}
P(S)^{IJ} = \int_S P^{IJ} .
\end{equation}
If $S$ is a 2-surface and $\ell$ is an edge `above' $S$, with source
node $n$ in $S$, being careful with numerical factors, one finds the
Poisson bracket between $P(S)^{IJ}$ and $U_{\ell}(n)$ to be exactly
\begin{equation}
\{P(S)^{IJ},U_{\ell}(n)\} =  \frac{1}{4} U_{\ell}(n) J^{IJ}
\end{equation}
where $(J^{IJ})^M{}_N:= 2 \delta^{M[I}\delta^{J]}_N$ is the basis
(modulo the skew symmetry in the $IJ$ label) of $\so(4)$ introduced
in appendix \ref{alg_app}.

Now, consider a graph $\gamma$ with only 4-valent nodes. Consider
the abstract triangulation $\Delta_3 = \gamma^*$ dual, within the
$3$-slice $\Sigma$, to the graph $\gamma$.  In the following, we
will motivate, by algebraic considerations, an identification, of
the the variables of the canonical discrete theory based on
$\Delta_3$, with certain variables of the present continuum theory.
Let $\Psi(\vec{U})$ be a function of the parallel transports along
the links in $\gamma$. Consider a node $n$ and adjacent link $\ell$
in $\gamma$.  Call the other three links at $n$ $\ell_1, \ell_2,
\ell_3$. Finally, let $t , f , f_1, f_2, f_3$ denote the tetrahedron
and abstract triangles in $\Delta_3$ dual to $n, \ell, \ell_1,
\ell_2, \ell_3$ respectively. Construct a 2-surface $S_{n,\ell}$
which intersects $\gamma$ only at the node $n$, with $\ell$ `above'
and $\ell_1, \ell_2, \ell_3$ `below' $S_{n,\ell}$.
%
%
The Poisson bracket of $P(S_{n,\ell})^{IJ}$ with $\Psi(\vec{U_l})$
is given by
\begin{equation}
\label{cyl_PB} \{P(S_{n,\ell})^{IJ},\Psi(\vec{U})\} =
\frac{1}{4}\left(\vec{L}_{(n,\ell)}^{IJ} - \sum_{i=1}^3
\vec{L}_{(n,\ell_i)}^{IJ}\right)\Psi(\vec{U})
\end{equation}
where $\vec{L}_{(n,\ell)}^{IJ}$ denotes the left invariant vector
field, on the parallel transport $U_{\ell}(n)$, associated to the lie
algebra element $J^{IJ}$.
Comparison of \Ref{cyl_PB} with the
discrete theory leads us to identify the following quantities in the
continuum and discrete theories
\begin{equation}
P(S_{n,\ell})^{IJ} = \frac{1}{4}\left(J_{f}(t)^{IJ}
- \sum_{i=1}^3 J_{f_i}(t)^{IJ}\right)
\end{equation}
where, as defined in section \ref{modelcl_sect}, $J_{\ell}(n)^{IJ}$ denotes the phase
space function in the discrete theory whose Hamiltonian vector field
is the left-invariant vector field on $U_{\ell}(n)$ corresponding to
the Lie algebra element $J^{IJ}$. Imposing the closure constraint in
the discrete theory reduces this to
\begin{equation}
P(S_{n,\ell})^{IJ} = \frac{1}{2} J_{f}(t)^{IJ} .
\end{equation}
Now if we define $B(S)^{IJ} := \int_S B^{IJ}$,
we have
\begin{eqnarray}
\nonumber
B(S_{(n,\ell)})^{IJ} &=& 2\Gfactor\left(\frac{\gamma^2}{\gamma^2-1}\right)
\left(P(S_{(n,\ell)})^{IJ} -
\frac{1}{\gamma} \dual P(S_{(n,\ell)})^{IJ}\right) \\
\nonumber
&=& \Gfactor \left(\frac{\gamma^2}{\gamma^2-1}\right)
\left(J_{f}(t)^{IJ} -
\frac{1}{\gamma} \dual J_{f}(t)^{IJ}\right) \\
\label{b_eqn}
&=& \Gfactor \lambda B_{f}(t)^{IJ}
\end{eqnarray}
where in the last line we have inserted \Ref{scaled_jb}. Recall the
physical meaning of $B_{f}(t)^{IJ}$: it is the integral of the two
form $B = \dual e \wedge e$ over the triangle $f$ dual to $\ell$. As
the triangulation $\Delta_3$ is abstract, a priori this triangle $f$
is some abstract surface dual to $\ell$. The above equation
\Ref{b_eqn}, forced on us from algebraic considerations, motivates
us to identify this dual surface with $S_{(\ell,n)}$, so that
$B_{f}(t)^{IJ}$ is physically the same quantity as
$B(S_{(n,\ell)})^{IJ}$.  We thus conclude $\lambda = 1/\Gfactor$,
giving us the final relation between $J_{f}(t)^{IJ}$ and
$B_{f}(t)^{IJ}$
\begin{equation}
J_f(t) = \frac{1}{\Gfactor} \left(B_f(t) + \frac{1}{\gamma}\dual B_f(t)\right).
\end{equation}


\begin{thebibliography}{99}

\bibitem{rovellibook}
Rovelli C 2004 \textit{Quantum Gravity} (Cambridge: Cambridge UP)

\bibitem{alrev} Ashtekar A and Lewandowski J 2004
Background independent quantum gravity: a status report
\textit{Class. Quant. Grav.} \textbf{20} R53-R152


\bibitem{thiemannbook}
Thiemann T 2007 \textit{Modern Canonical Quantum General Relativity}
(Cambridge: Cambridge UP)

\bibitem{thomas}
Thiemann T 1998 Quantum spin dynamics (QSD)
\textit{Class. Quant. Grav.} \textbf{15} 893-873 \\
Thiemann T 2006 Quantum spin dynamics VIII. The Master Constraint
\textit{Class. Quant. Grav.} \textbf{23} 2249-2266

\bibitem{spinfoams}
Perez A 2003 Spin Foam Models for Quantum Gravity
\emph{Class. Quant. Grav.} \textbf{20} (2003) R43 \\
%
Oriti D 2001 Spacetime geometry from algebra: spin foam models for
non-perturbative quantum gravity \textit{Rept. Prog. Phys.}
\textbf{64} 1489-1544

\bibitem{bc}
Barrett J W and Crane L 1998 Relativistic spin networks and quantum
gravity
 \textit{J. Math. Phys.} \textbf{39} 3296-3302 \\
DePietri R, Freidel L, Krasnov K, and Rovelli C 2000 Barrett-Crane
model from a Boulatov-Ooguri field theory over a homogeneous space
\textit{Nucl.  Phys. } B\textbf{574} 785-806 \\
Perez A and Rovelli C 2001 A spinfoam model without bubble
divergences
\textit{Nucl.  Phys. } B\textbf{599} 255-282 \\
Oriti D and Williams R M 2001 Gluing 4-simplices: a derivation of
the Barrett-Crane spinfoam model for Euclidean quantum gravity
\textit{Phys.  Rev. } D\textbf{63} 024022

\bibitem{emanuele}
Alesci E and Rovelli C 2007 The full LQG graviton propagator: I.
Difficulties with the Barrett-Crane vertex, \textit{Preprint}:
\texttt{arXiv:0708.0883}

\bibitem{eprlett}
Engle J, Pereira R, and Rovelli C 2007 Loop-quantum-gravity
vertex-amplitude \textit{Phys. Rev. Lett.} \textbf{99} 161301

\bibitem{eprpap}
Engle J, Pereira R, and Rovelli C 2007 Flipped spinfoam vertex and
loop gravity, \textit{Preprint}: \texttt{arXiv:0708.1236}

\bibitem{ls_cohstates}
Livine E and Speziale S 2007 A new spinfoam vertex for quantum
gravity, \textit{Preprint}: \texttt{arXiv:0705.0674}

\bibitem{fk2007}
Freidel L and Krasnov K 2007 A new spin foam model for 4d gravity,
\textit{Preprint}: \texttt{arXiv:0708.1595}

\bibitem{ls_model}
Livine E and Speziale S 2007 Consistently solving the simplicity
constraints for spinfoam quantum gravity, \textit{Preprint}:
\texttt{arXiv:0708.1915}

\bibitem{gengamma}
Engle J, Livine E, Pereira R, and Rovelli C 2007 LQG vertex with
finite Immirzi parameter. \textit{Preprint}:
\texttt{arXiv:0711.0146}

\bibitem{master}
Thiemann T 2006 The Pheonix project: master constraint programme for
loop quantum gravity \textit{Class. Quant. Grav.} \textbf{23} 4453-4472 \\
Dittrich B and Thiemann T 2006 Testing the master constraint
programme for loop quantum gravity I. General framework
\textit{Class. Quant. Grav.} \textbf{23} 1025-1066 \\
Dittrich B and Thiemann T 2006 Testing the master constraint
programme for loop quantum gravity II. Finite dimensional systems
\textit{Class. Quant. Grav.} \textbf{23} 1067-1088 \\
Dittrich B and Thiemann T 2006 Testing the master constraint
programme for loop quantum gravity III. SL(2,R) models
\textit{Class. Quant. Grav.} \textbf{23} 1089-1120

%

%
%

\bibitem{holst} Holst S 1996 Barbero's Hamiltonian derived from a
generalized Hilbert-Palatini action \textit{Phys. Rev.} D\textbf{53}
5966-5969

\bibitem{eteraBF}
Livine E and Oriti D 2002 Barrett-Crane spin foam model from
generalized BF type action for gravity {\it Phys. Rev.} D\textbf{65}
044025

\bibitem{eteraprojsn}
Livine E 2002 Projected spin networks for Lorentz connection:
Linking spin foams and loop gravity {\it Class. Quant. Grav.} {\bf
19} 5525-5542

\bibitem{reisenberger}
Reisenberger M 1999 On relativistic spin network vertices
\textit{J. Math. Phys.} \textbf{40} 2046-2054

\bibitem{engle2006}
Engle J 2006 Quantum field theory and its symmetry reduction
\textit{Class. Quant. Grav.} \textbf{23} 2861-2894

\bibitem{asht_tate}
Ashtekar A, Tate R 1994 An algebraic extension of Dirac
quantization: Examples \textit{J. Math. Phys.} \textbf{35} 6434-6470

%
%

\bibitem{sergei2007}
Alexandrov S 2007 Spin foam model from canonical quantization,
\textit{Preprint}: \texttt{arXiv:0705.3892}

\bibitem{carlospec}
Rovelli C 2007 Comment on ``Are the spectra of geometrical operators
in Loop Quantum Gravity really discrete?'' by B. Dittrich and T.
Thiemann, \texttt{arXiv:0708.2481}

\bibitem{partialobs}
Rovelli C 2002 Partial observables \textit{Phys. Rev.} D\textbf{65}
124013

\bibitem{flippedlor}
Pereira R 2007 Lorentzian LQG vertex amplitude. 
\textit{Preprint}: \texttt{arXiv:0710.5043}

\bibitem{perelomov}
Perelomov AM, \emph{Generalized Coherent States and Their
Applications}(Springer-Verlag, 1986)

\bibitem{hartlesorkin}
Hartle J B and Sorkin R 1981 Boundary terms in the action for the
Regge calculus {\it Gen. Rel. Grav.} {\bf 13} 541-549

%
%
\bibitem{asht_et}
A Ashtekar, L Bombelli and O Reula ``Covariant phase space of
asymptotically flat gravitational fields'', in \textit{Mechanics,
Analysis and Geometry: 200 Years after Lagrange}, Ed. by M
Francaviglia and D Holm (North Holland, Amsterdam 1991). \\
A Ashtekar, L Bombelli, and R Koul ``Phase space formulation of
General Relativity without 3+1 splitting'', in \textit{The Physics
of Phase Space}, Ed. by Y S Kim and W W Zachary (Springer-Verlag,
Berlin, 1987).

%
%

\end{thebibliography}
\end{document}